\documentclass[preprint,prd,showpacs,showkeys,preprintnumbers,amsmath,amssymb]{revtex4}

\newcommand{\beq}{\begin{equation}}
\newcommand{\eeq}{\end{equation}}

\newcommand{\bfxi}{\mbox{\boldmath $\xi$}}

\newcommand{\bfx}{{\bf x}}

\begin{document}

\preprint{APS/123-QED}

\title{Weak field limit of Reissner-Nordstr\"{o}m black hole lensing}

\author{Mauro Sereno}
\email{sereno@na.infn.it}
\affiliation{Dipartimento di Fisica ``E.R. Caianiello", \\
Universit\`{a} di Salerno, Via S. Allende, 84081 Baronissi (Salerno),
Italia}

\date{July 4, 2003}

\begin{abstract}
We study gravitational lensing by a Reissner-Nordstr\"{o}m (RN) black hole
in the weak field limit. We obtain the basic equations for the
deflection angle and time delay and find analytical expressions for
the positions and amplifications of the primary and secondary images.
Due to a net positive charge, the separation between images increases,
but no change in the total magnification occurs.
\end{abstract}

\maketitle

\section{Introduction}

Within the next decade, the accuracy of space-based missions dedicated
to measuring astrometric positions, parallaxes, and proper motions of
stars is expected to attain 1 microarcsecond ($\mu$as). Thus progress
in observational techniques has made it necessary to take into account
many subtle relativistic effects in light propagation. Lensing by
massive black holes requires the full relativistic corrections,
demanding a treatment of lensing theory to the proper order of
approximation at which interesting effects appear for any particular
phenomenon.

The gravitational lensing can reveal an unique tool to detect exotic
objects in the universe, that, though not yet observed, are not
forbidden on a theoretical ground. In addition to the primary and
secondary images, a spherically symmetric black hole produces two
infinite series of faint relativistic images, formed by light rays
winding around the black hole at distances comparable to the
gravitational radius. The weak field theory of gravitational lensing
suffices to describe the primary and secondary images, but strong
field lensing is demanded to consider phenomena close to the horizon.
Since the relativistic images are strongly demagnified and the
astrometric separations among all the images are really small, a
proper consideration of the full system of images requires a treatment
to higher orders of approximation of the images formed in the weak
field regime.

In the present work, we wish to study the lensing situation then the
deflector is a charged or Reissner-Nordstr\"{o}m (RN) black hole. Charged
rotating black holes are plausible endpoints of the catastrophic
gravitational collapse of the most massive magnetized rotating stars.
When surrounded by a co-rotating magnetosphere of equal and opposite
charge, the system attains a minimum energy configuration. The
magnetosphere preserves the black hole from a neutralization due to a
selective accretion of charge from the environment and the black hole
can be quite stable in a typical astrophysical environment of low
density \cite{pun98}.

In this paper, we propose a comprehensive analysis of the weak field
limit of gravitational lensing due to a charged black hole.
Gravitational lensing in the strong field scenario when the lens is a
RN black hole has been discussed in Eiroa et al. \cite{eir+al02}, who
argued that rotation does not lead to qualitatively different results.
They found analytical expressions for the positions and amplifications
of the two sets of relativistic images, but they did not discuss the
effect of a charge on the primary and secondary images.

Bhadra \cite{bha03} discussed gravitational lensing due to a charged
black hole of heterotic string theory with the aim of examining the
possible string effects in a strong-field observation, but no
significant signatures emerged in lensing observables.

In Section~\ref{rnweak}, we perform the weak field limit of the RN
metric. In Section~\ref{rnlens}, following Fermat's principle, we
calculate analytical expressions for both time delay and deflection
angle. In Section~\ref{rnmap}, using a perturbative method, we
calculate the positions and magnification of the images.
Section~\ref{rnconc} is devoted to some final considerations. In this
paper, we will use geometrized units (speed of light in vacuum $c=1$
and gravitational constant $G=1$).

\section{The RN metric}
\label{rnweak}

The RN metric is a spherically symmetric solution of the coupled
equations of Einstein and of Maxwell. It represents a black hole with
a mass $M$ and a charge $Q$. The RN metric, in its standard form,
reads
\begin{equation}
\label{rn1}
ds^2=\left(1-\frac{2M}{r}+\frac{Q^2}{r^2} \right)dt^2 -
\left(1-\frac{2M}{r}+\frac{Q^2}{r^2}
\right)^{-1} dr^2 -r^2 \left( \sin^2 \theta d \phi^2 + d
\theta^2 \right).
\end{equation}
The metric in Eq.~(\ref{rn1}) can be expressed in an equivalent
isotropic form by introducing a new radius variable, $\rho$,
\begin{equation}
\label{rn2}
 r = \rho \left( 1+ \frac{M}{\rho} +  \frac{M^2 - Q^2}{4 \rho^2} \right),
\end{equation}
Substituting Eq.~(\ref{rn2}) in Eq.~(\ref{rn1}) gives
\begin{eqnarray}
\label{rn3}
ds^2 & = & \left[ \frac{M^2-4\rho^2-Q^2}{(M+2\rho)^2-Q^2)} \right]^2
dt^2 \\ & -&
\left( 1+\frac{M}{\rho} + \frac{M^2-Q^2}{4\rho^2} \right)^{2}
\left\{ d \rho^2 +\rho^2 \left( \sin^2 \theta d \phi^2 + d \theta^2 \right) \right\}. \nonumber
\end{eqnarray}
Finally, in quasi-Minkovskian coordinates, the weak field limit of
Eq.~(\ref{rn3}) reads
\begin{equation}
\label{rn4}
ds^2  \simeq  \left[ 1-\frac{2M}{x}+\left( 1+ \frac{a^2}{2}\right)
\frac{2M^2}{x^2}
\right] dt^2 -\left[ 1+\frac{2M}{x}+\left( 1- \frac{a^2}{3}\right)
\frac{3}{2}\frac{M^2}{x^2} \right]
d \bfx^2,
\end{equation}
where $a \equiv Q/M \leq 1$.

\section{Lensing quantities}
\label{rnlens}

The time delay of the deflected path $p$ relative to the unlensed ray
$p_0$ is
\begin{equation}
\label{wf6}
\Delta T \equiv  \int_p n \ dl_{\rm P} - \int_{p_0} dl_{\rm P},
\end{equation}
being $n$ the effective index of refraction, defined as
\begin{equation}
\label{staz4}
n \equiv -\frac{g_{i0}}{g_{00}} e^i+\frac{1}{\sqrt{g_{00}}},
\end{equation}
where $e^i \equiv \frac{dx^i}{d l_{\rm P}}$ is the unit tangent vector
of a ray; ${d l_{\rm P}}^2 \equiv \left( -g_{ij}
+\frac{g_{0i}g_{0j}}{g_{00}}\right)dx^i d x^j$ defines the spatial
metric.

For the RN metric, we get
\begin{equation}
d l_{\rm P} \simeq \left\{ 1+ \frac{M}{x} +
\frac{1-a^2}{4} \frac{M^2}{x^2} \right\} d l_{\rm eucl},
\end{equation}
where $d l_{\rm eucl} \equiv \sqrt{ \delta_{ij}d x^i d x^j}$ is the
Euclidean arc length, and
\begin{equation}
n \simeq 1+ \frac{M}{x} +
\frac{1-a^2}{2}\frac{M^2}{x^2}.
\end{equation}

Equation~(\ref{wf6}) can be expressed as a sum of geometrical and
potential time delays
\[
\Delta T =\Delta T_{\rm geom}+\Delta T_{\rm pot}.
\]
The geometrical time delay,
\begin{equation}
\label{wf7}
\Delta T_{\rm geom} \equiv \int_p \ d l_{\rm P} - \int_{p_0} \ d
l_{\rm P},
\end{equation}
is due to the extra path length relative to the unperturbed ray $p_0$.

It is useful to employ the spatial orthogonal coordinates $(\xi_1,
\xi_2, l )$, centred on the lens and such that the $l$-axis is along
the incoming light ray direction ${\bf e}_{\rm in}$. The
three-dimensional position vector $\bf x$ can be expressed as ${\bf x}
= \mbox{\boldmath $\xi$} + l {\bf e}_{\rm in}$. The lens plane,
$(\xi_1,\xi_2)$, corresponds to $l=0$. With these assumptions, the
geometrical time delay is \cite{sef,pet+al01}
\begin{equation}
\label{wf9}
\Delta T_{\rm geom} \simeq \frac{1}{2}\frac{D_{\rm d} D_{s}}{D_{\rm
ds}}\left|
\frac{\mbox{\boldmath $\xi$}}{D_{\rm d}}-\frac{\mbox{\boldmath $\eta$}}{D_{\rm s}} \right|^2,
\end{equation}
where $D_{\rm s}$ is the distance from the observer to the source,
$D_{\rm d}$ is the distance from the observer to the deflector and
$D_{\rm ds}$ is the distance from the deflector to the source;
$\mbox{\boldmath $\eta$}$ is the two-dimensional vector position of
the source in the source plane.

The potential time delay $\Delta T_{\rm pot}$ accounts for the
retardation of the deflected ray caused by the gravitational field of
the lens; it is defined as the difference between the total travel
time and the integral of the line element along the deflected path,
\begin{equation}
\Delta T_{\rm pot} \equiv \int_p \ n\ d l_{\rm P} - \int_{p} \ d
l_{\rm P}.
\end{equation}
To calculate properly the potential time delay, we have to consider
the time delay to the post-post-Newtonian (ppN) order \cite{io03prd}.
The first contribution derives from the integration of the Newtonian
potential on the deflected path, calculated at order $G$. Non-linear
interaction of matter with space-time, represented in the metric
element by terms which contain $1/x^2$, also contributes. Finally,
higher order corrections to the geometrical time delay enter the ppN
time delay, since the difference in the distances of closest approach
of the deflected and undeflected light rays represents a
post-Newtonian quantity and first-order corrections in the calculation
of this difference induce a ppN contribution to the time delay.

After lengthy but straightforward calculations, we get
\begin{equation}
\label{rn19}
\Delta T \simeq -4 M \ln \left( \frac{\xi}{\xi_0} \right)
+ \frac{3 \pi}{4} \left( 5 - a^2 \right) M^2\frac{1}{\xi},
\end{equation}
where $\xi_0$ is a scale-length in the lens plane and $\xi$ is the
impact parameter.

From the Fermat's principle \cite{sef,pet+al01}, we can derive the
deflection angle, i.e. the difference of the initial and final ray
direction,
\begin{equation}
\hat{\mbox{\boldmath $\alpha$}} \equiv -
\nabla_{\mbox{\boldmath $\xi$}}
\Delta T_{\rm pot}.
\end{equation}
We get
\begin{equation}
\label{rn20}
\hat{\mbox{\boldmath $\alpha$}} (\bfxi) \simeq 4M \frac{\bfxi}{\xi^2} +
\frac{3 \pi}{4} \left( 5 - a^2 \right) M^2\frac{\bfxi}{\xi^3}.
\end{equation}
The curvature of the space-time generated by the charge induces a
correction to the ppN order. Again, using the Fermat's principle, the
lens equation reads
\begin{equation}
\label{wf17}
\mbox{\boldmath $\eta$}=
\frac{D_{\rm s}}{D_{\rm d}}\mbox{\boldmath $\xi$}
-D_{\rm ds}\hat{\mbox{\boldmath $\alpha$}}(\mbox{\boldmath $\xi$}).
\end{equation}

\section{The lens mapping}
\label{rnmap}

Owing to the spherical symmetry of the system, the lens equation
reduces to a one-dimensional equation: source, lens and images lie on
a straight line on the observer's sky.

As a natural length scale, we introduce the Einstein radius $R_{\rm
E}$,
\begin{equation}
\xi_0= R_{\rm E} \equiv \sqrt{4M \frac{D_{\rm d} D_{\rm ds}}{D_{\rm s}}}.
\end{equation}. Let us change to a dimensionless variable $x \equiv
\xi /\xi_0$.
The lens equation reads
\begin{equation}
\label{lens1}
y = x - \alpha (x),
\end{equation}
where $y$ is the position of the source, in units of $\frac{D_{\rm
s}}{D_{\rm d}}
\xi_0$, and $\alpha$ is the scaled deflection angle,  $ \alpha (x) \equiv
\frac{D_{\rm d} D_{\rm ds}}{\xi_0 D_{\rm s}} \hat{\alpha}(\xi)$. Owing to
symmetry, we will consider source positions $ y \geq 0$, whereas the
range of $x$ will be taken to be the whole real axis.

The scaled deflection angle becomes
\begin{equation}
\alpha (x)= \frac{1}{x}+\frac{3\pi}{32}\left( 5-a^2 \right) \frac{R_{\rm Sch}}{R_{\rm E}}\frac{1}{x|x|}.
\end{equation}
In the next, we will use the notation $d_{\rm RN} \equiv
\frac{3\pi}{32}\left( 5-a^2 \right)
\frac{R_{\rm Sch}}{R_{\rm E}}$.

Under the condition $d_{\rm RN} \ll 1$, that holds in usual
astrophysical systems, we can perform a perturbative analysis to
obtain approximate solutions of the lens equation \cite{io03hom}. The
image position can be expressed as,
\begin{equation}
\label{pert}
x \simeq x_{(0)}+d_{\rm RN} x_{(1)},
\end{equation}
where $x_{(0)}$ and $x_{(1)}$ denote the zeroth-order solution and the
correction to the first-order, respectively.  The unperturbed images
are solutions of the lens equation for $d_{\rm RN}=0$. Two unperturbed
images form at
\begin{equation}
x^{\pm}_{(0)}=\frac{1}{2}\left( 1 \pm \sqrt{1 +\frac{4}{y^2}}\right) y .
\end{equation}
The primary image $x_+$ lies outside the Einstein ring (on the same
side of the source), while the secondary image $x_-$ is inside (on the
side opposite the source).

By substituting the expression in Eq.~(\ref{pert}) for the perturbed
images in the full lens equation, Eq.~(\ref{lens1}), we obtain the
first-order perturbations,
\begin{equation}
x^{\pm}_{(1)} =  x^{\pm}_{(0)} +\frac{d_{\rm RN}}{x^{\pm}_{(0)}\sqrt{y^2+4}},
\end{equation}

Critical curves form where the Jacobian of the lens mapping vanishes.
It is
\begin{equation}
A=\frac{y}{x}\frac{dy}{dx} \simeq 1
-\frac{1}{x^4}-\frac{3-x^2}{|x|^5}d_{\rm RN}.
\end{equation}
$A$ is singular at the tangential critical curve, a circle centred on
the origin and mapped onto the point $y=0$. The effect of a charge is
to increase the radius of the tangential circle from 1 to
$1+\frac{d_{\rm RN}}{2}$. No critical radial curve forms.

The amplification $\mu = 1/|A|$ for each image becomes
\begin{equation}
\mu^{\pm}=\frac{y^2+2}{2y\sqrt{y^2+4}} \pm \left( \frac{1}{2} - \frac{d_{\rm RN}}{(y^2+4)^{3/2}} \right).
\end{equation}
As noted in \cite{ebi+al00}, a correction $\propto 1/x^2$ to the
deflection angle does not change the total magnification of a source
at first order. Only the amplification factor for each image is
corrected. Furthermore, no effects in astrometric shift during
microlensing events are produced \cite{le+wa01}.

\section{Conclusions}
\label{rnconc}

We have discussed gravitational lensing due to a RN black hole in the
weak field limit. A charge in the lens induces a correction of ppN
order in the time delay of images and deflection angle. We found the
position and magnification of the images. Effects of the charge occur
only for the quantities relative to each image whereas the total
amplification curve and the astrometric shift in microlensing events
are not affected.

Despite we have developed our lensing formalism for a RN black hole,
our procedure is quite general and can be extended to any
astrophysical system wherein a correction of ppN order to the lensing
quantities arises in the weak field limit. Only the form of the
parameter $d_{\rm RN}$ changes as a function of the parameters that
characterizes the system.

In view of using gravitational lensing either to discover exotic
objects in the universe or to test alternative theories of gravity, a
methodical study of higher-order effects is demanded. Virbhadra et
al.~\cite{vir+al98} investigated a static and circularly symmetric
lens with mass and scalar charge. Such an object can be also treated
as a naked singularity. Due to the scalar field, an extra term
$1/\xi^2$ appears.

Gravitational lensing in metric theories of gravity can also reproduce
some feature of the RN lensing. Such theories can be characterized by
an approximate metric element, with dimensionless parameters, derived
in the slow-motion and weak field limit, inclusive of both ppN
corrections and gravito-magnetic field \cite{io03prd}. When the
standard coefficients of the post-Newtonian parametrized expansion
reduce to $\beta
=1+\frac{a^2}{2}$ and $\gamma =1$, the ppN coefficient
$\epsilon$ to $1-\frac{a^2}{3}$ and the gravito-magnetic one, $\mu$,
to zero, then the metric tensor reduces to the weak field limit of the
RN metric. Using these particular values for the coefficients in the
formulae derived for time delay and deflection angle in metric
theories of gravity \cite{io03prd}, we retrieve the expression in
Eqs.~(\ref{rn19},~\ref{rn20}).

Future astrometric space missions with a resolution of $\mu$as, such
as the ARISE project \cite{ulv99}, demands a full knowledge of the
phenomenology of the primary and secondary images produced in the weak
field limit. This is a fundamental step in the detection of
relativistic images of charged black hole, that are closed to the
nonrelativistic ones but are much less intense \cite{eir+al02}.

\end{document}